\documentclass[epjCONF,columns]{svjour} 
\usepackage{graphics}
\usepackage[varg]{txfonts} 
\usepackage[latin1]{inputenc}

\newcommand{\beq}{\begin{equation}}
\newcommand{\eeq}{\end{equation}}
\newcommand{\bea}{\begin{eqnarray}}
\newcommand{\eea}{\end{eqnarray}}

\session-title{International Conference Nuclear Structure and Related Topics Dubna, Russia, July 2 - July 7, 2012}
\begin{document}
\title{Quadrupole moments of odd-odd near-magic nuclei}
\author{D. Voitenkov \inst{1}\fnmsep\thanks{\email{voitenkov@list.ru}} \and O. Achakovskiy \inst{2} \and S. Kamerdzhiev \inst{1} \fnmsep \inst{2} \and  S. Tolokonnikov \inst{3}}
\institute{Institute for Physics and Power Engineering, 249033 Obninsk, Russia 
\and Insitute for Nuclear Power Engineering, National Research Nuclear University MEPhI, 249040 Obninsk, Russia 
\and Kurchatov Institute, 123182 Moscow, Russia}
\abstract{
Ground state quadrupole moments of odd-odd near double magic nuclei are calculated in the approximation 
of  no interaction between odd particles. Under such a simple approximation, the problem is reduced to the calculations of quadrupole moments of corresponding odd-even nuclei. These calculations are performed  within the self-consistent Theory of Finite Fermi Systems based on the Energy Density Functional by Fayans et al. with the known  DF3-a parameters.  A reasonable agreement with the available experimental data has been obtained for odd-odd nuclei and odd near-magic nuclei investigated. The self-consistent approach under consideration allowed us to predict the unknown quadrupole moments of odd-even and odd-odd nuclei near  the double-magic $^{56,78}$Ni, $^{100,132}$Sn ones.
} 
\maketitle
\section{Introduction}
\label{intro}
One of the challenging goals of modern nuclear theory is to elaborate the approaches which would 
enable the characteristics of unstable nuclei to be predicted. 
Therefore, a developed theoretical approach should have a very high  predictive power. The Energy Density Functional (EDF)  approach is one of  such  approaches.
As a rule, odd-odd nuclei are unstable ones, so here  we will consider their characteristics using the self-consistent Theory of Finite Fermi Systems (TFFS) \cite{khodelsap} based on the EDF by Fayans et al.\cite{Fay}. 

Quite  recently, a good description of the ground state quadrupole \cite{BE2,QM} and magnetic\cite{Tol-Sap} moments of odd near- and semi-magic nuclei has been obtained within this self-consistent approach  where the ``single quasi-particle approximation'' developed in the framework of the standard TFFS \cite{AB} has been used. This approximation means that one 
quasi-particle with the local charge $e_q$ is added to the even-even core and  the  core is polarized due to the  Landau-Migdal (LM) interaction between  the particle considered and the core nucleons. In other words, the quasi-particle possesses the effective charge $e_{eff}$ caused by the polarizability of the core, which should be found by solving the TFFS equations. Within the TFFS, static moments of odd nucleus are determined by the 
diagonal matrix element of the effective field $V$ in the external field  $V_{0}$ (see  Eq.(\ref{Vef_s}) below).

As the odd-odd nuclei are more complicated objects than the even-odd ones, here we only consider  the odd-odd near-magic
nuclei  together with the corresponding odd near-magic one. Within the above-mentioned self-consistent EDF approach, we will calculate  the ground state quadrupole moments of odd-odd near-magic nuclei    with the use, for simplicity, of the approximation that there is no interaction between two odd particles. As we will see,  such a simple approximation 
allows us to check this approximation rather successfully in a pure phenomenological way. 
 
\section{Quadrupole moments of odd and odd-odd near-magic nuclei}

Within  the TFFS \cite{AB} , the static quadrupole moment $Q_{\lambda}$ of an
odd near-magic nucleus with the  odd nucleon in the state $\lambda \equiv (n, j, l, m)$  can be
found in terms of the diagonal matrix element 
 of the effective field $V$: 
  
\beq\label{quodd}
 Q_\lambda = \langle\lambda|
V(\omega =0)|\lambda\rangle = c_{j} \langle njl \parallel V \parallel njl \rangle , 
\eeq
where 3j-symbol $c_{j}=2j(2j-1)^{1/2}\left[(2j+3)(2j+2)\right. \times $ \\$\left.  \times(2j+1)2j \right]^{-1/2}$.
In the framework of the above-mentioned TFFS single quasi-particle approximation and  of our main approximation 
that there is  no interaction between two quasi-particles, the quadrupole moment of odd-odd nucleus with spin $I$ is as follows  

\beq\label{qu}
Q_I=<II \mid V^p + V^n \mid II>,
\eeq
where $\Psi_{II} = \Sigma \varphi_1 \varphi_2 <j_1m_1 j_2m_2\mid II>$, for the case of odd particle-odd particle. 
Here  $\varphi_1$ is the single-particle wave function with the quantum numbers 
$1 \equiv \lambda_1 \equiv (n_1,j_1,l_1,m_1)$. In the limit of no interaction between the core and  odd quasi-particles
the effective fields $V^p$ and $V^n$ should be replaced  by $e_{q}^{p}V_0$ and $e_{q}^{n}V_0$, see Eq. (\ref{Vef_s})
below.

Then the expression for ground state quadrupole moments of odd-odd near-magic nuclei has the form (for the case of particle-particle or hole-hole):
\bea\label{Qoddodd}
 Q_I = (2I+1)
\left(\begin{array}{ccc}
{I} &{2} &{I}\\
{I} &{0} &{-I} \end{array}\right) 
(-1)^{j_p+j_n+I+2} \times\nonumber\\
\times \left[\left\lbrace\ \begin{array}{ccc}
{j_p} &{I} &{j_n}\\
{I} &{j_p} &{2}
\end{array} \right\rbrace c^{-1}_{j_{p}} Q^{p} 
+ \left\lbrace \begin{array}{ccc}
j_n & I & j_p\\
I & j_n & 2
\end{array}\right\rbrace  c^{-1}_{j_{n}} Q^{n}  \right] ,
\eea
 and for the case of particle-hole we have 
 \bea\label{Qoddoddph}
  Q_I = (2I+1)
\left(\begin{array}{ccc}
{I} &{2} &{I}\\
{I} &{0} &{-I} \end{array}\right) 
(-1)^{j_p+j_n+I+2}\times \nonumber\\
\times \left[\left\lbrace\ \begin{array}{ccc}
{j_p} &{I} &{j_n}\\
{I} &{j_p} &{2}
\end{array} \right\rbrace c^{-1}_{j_{p}} Q^{p} 
+(-1)^{j_p-j_n} \left\lbrace \begin{array}{ccc}
j_n & I & j_p\\
I & j_n & 2
\end{array}\right\rbrace  c^{-1}_{j_{n}} Q^{n}  \right] ,
 \eea
 where $Q^p$ and $Q^n$ are the quadrupole moments of corresponding odd nuclei which are determined by Eq.(\ref{quodd}). A similar formula can be easily obtained for the hole-particle case.

Thus, within such a simple approximation, the problem is reduced to the calculations of quadrupole moments of corresponding odd-even nuclei. One can take the values of quadrupole moments of odd nuclei from the experiment (phenomenological approach, Sect. \ref{phenapp}) and   one can also calculate them and  obtain  the quadrupole moments of corresponding odd-odd nuclei according to Eqs. (3,4) (microscopic approach, Sects. \ref{selfcalc})

\section{Phenomenological approach}
\label{phenapp}
We have found only three odd-odd nuclei where there are all three necessary experimental values.
Our main approximation  is confirmed by a reasonable agreement with the experiment in Table 1, where the experimental values of quadrupole moments of corresponding odd nuclei have been used. The  experimental data used here and  what follows have been taken from \cite{stone}.
\begin{table}[th]
\caption{Quadrupole moments \textit{Q} (e b) of odd-odd near-magic nuclei (phenomenological approach).}
\label{tablphen}
\begin{center}
\begin{tabular}{l c c c c }
\hline \hline \noalign{\smallskip}
 nucl.  & $J^{\rm \pi}$
&\hspace*{1.ex} $T_{\rm 1/2}$\hspace*{1.ex} &\hspace*{1.ex} $Q_{\rm phen}$ &\hspace*{1.ex}$Q_{\rm exp}$\\
\noalign{\smallskip}
\hline
\noalign{\smallskip}
${^{40}_{19}}$K$_{21}$ & 4$^{-}$    & 1.248$\times$10$^{9}$ y   & -0.106(6)  & -0.071(1) \\
${^{92}_{41}}$Nb$_{51}$ & 7$^{+}$   & 3.47$\times$10$^{7}$  y   & -0.43(7)   & -0.35(3)\\
${^{210}_{83}}$Bi$_{127}$ & 1$^{-}$ & 5.01           d   & +0.22(6)   & +0.19(6)\\
\\
\hline \hline
\end{tabular}
\end{center}
\end{table}

More convincing  results which confirm our approximation have been obtained for static ground state magnetic moments of odd-odd near-magic nuclei  where there are much more experimental data, see 
Ref. \cite{kaevyadfiz}

\section{Self-consistent calculations}
\label{selfcalc}
These calculations are performed within the self-consistent Theory of Finite Fermi Systems (TFFS) based on the Energy Density Functional (EDF) by Fayans et al. with the known DF3-a parameters  \cite{Tol-Sap1}.
The details of the calculations  for odd nuclei have been described in Ref. \cite{BE2}. Here we summarize
several formulas which are required for understanding the main
ingredients of the approach. 
In this method, the ground
state energy of a nucleus is considered as a functional of normal
and anomalous densities, 
\beq E_0=\int {\cal E}[\rho_n({\bf
r}),\rho_p({\bf r}),\nu_n({\bf r}),\nu_p({\bf r})] d^3r.\label{E0}
\eeq

According to Eq.(1), the static quadrupole moment $Q_{\lambda}$ of an
odd nucleus with the  odd nucleon in the state $\lambda$  can be
found by solving  the equation 
 for the effective field $V$ in the
static external field 
$ V_0({\bf r}) = 
\sqrt{\frac{16 \pi} {5}} r^2 Y_{20}({\bf n})$. 
\beq
\label{Vef_s}{\hat V}(\omega)= {\hat e_q} V_0(\omega)+{\hat {\cal F}}
{\hat A}(\omega) {\hat V}(\omega), 
\eeq 
where $e_q$ is the local
quasi-particle charge with respect to the external field $V_0$
 and all  terms  are matrices in the isotopic space. 
In the standard TFFS notation, the particle-hole propagator $A$
in the coordinate representation reads: 

\bea\label{A}
A(\textbf{r},\textbf{r}^\prime, \omega ) = -\sum_{1} n_1 \varphi_{1}^{*}(\textbf{r}) \varphi_1(\textbf{r}^\prime )\times \nonumber\\ 
\times\left[G(\textbf{r}^\prime ,\textbf{r}; \epsilon _1 + \omega ) +  G(\textbf{r}^\prime ,\textbf{r}; \epsilon _1 -\omega )\right] 
\eea
where $n_1 = (0,1)$ are occupation numbers, summation is over states below the Fermi surface
and the known single-particle Green functions $G$ already contain  the entire single-particle continuum.
Just due to this feature of the Green function, the single-particle continuum is taken into account completely in the TFFS.

 In our case, the local charges in Eq. (\ref{Vef_s}) are $e_q^p=1,\; e_q^n=0$.
The explicit form of the above equations
 is written down for the case of the electric ($t$-even)
symmetry we deal with.
 In Eq. (\ref{Vef_s}), ${\cal F}$ is the usual LM amplitude,
\beq {\cal F}=\frac {\delta^2 {\cal E}}{\delta \rho^2},
\label{LM}\eeq

Using formulas (\ref{E0}) -- (\ref{LM}) we have calculated self-consistently ground state quadrupole moments of odd and corresponding odd-odd near-magic nuclei. The method used has a high predictive power and, according to Eq. (7), takes into account completely the single-particle continuum,which is especially important for nuclei with the small 
separation energy, like $^{78}$Ni ($S_n \approx$ 5.8 MeV) and $^{100}$Sn ($S_p \approx$ 2.4 MeV) . For this reason, first of all we have calculated the quadrupole moments values of the odd and odd-odd nuclei near unstable double-magic nuclei $^{56}$Ni, $^{78}$Ni, $^{100}$Sn and $^{132}$Sn.      

\subsection{Odd near-magic nuclei}
\label{selfodd}
The final expression for the quadrupole moment of an odd nucleus, Eq.(1),
is as follows

 \beq\label{Vlam}
 Q_{\lambda} = V_{\lambda} =  \pm \frac{2j-1}{2j+2} \int  V(r) R_{nlj}^2(r)
 r^{2}dr.
 \eeq
  where the - sign  should be taken  for the odd particle and + stands for the odd hole  \cite{BM}.
  (The replacement of the effective field $V$ by $e_{q}V_{0}$ which was considered in \cite{BM}
  does not change the sign).  
 The $j$-dependent factor in (\ref{Vlam})
appears due to the angular integral. For $j>1/2$ it is
always negative. The equation for the quantity $V(r)$, which  was obtained from Eq.(\ref{Vef_s}),
has been solved in the coordinate space using Eqs.(\ref{A},\ref{LM}). The same set of DF3-a parameters 
has been used to calculate the self-consistent single-particle basis and the effective LM interaction.  

The results of calculations are shown in Table 2 and Table 3, where in column Q$_{\rm theor}$ the self-consistent values  are shown for odd nuclei near double-magic nuclei $^{56,78}$Ni, $^{100,132}$Sn, $^{208}$Pb. To compare with the  well-known phenomenological description, the results with the effective charger $e{^p_{eff}}=2$, $e{^n_{eff}}=1$  are represented in columns Q${\rm _{eff}}$, where they have been calculated with the same self-consistent single-particle basis. These values have been justified microscopically within TFFS 
in Ref. \cite{kaevyadfiz1965} where they have been introduced as 
$(e^p_{eff})_{\lambda} =  V_{\lambda}^p$/$(V_{0})_{\lambda}$ and 
 $e{^n_{eff}} = V_{\lambda}^n$/$(V_{0})_{\lambda}$. Note that in nuclei with paring these values are $e_{eff} \cong 4-6$ \cite{BE2} due to the contribution of unfilled nuclear shells. Comparing the values in columns $Q_{eff}$ and $Q_{theor}$,
 one can see that the use of the phenomenological effective charger $e{^p_{eff}}=2$, $e{^n_{eff}}=1$ is not always good quantitatively. 
 
\begin{table}[th]
\label{tablqoddn}
\caption{Quadrupole moments \textit{Q} (e b) of odd-neutron near-magic nuclei.}
\begin{center}
\begin{tabular}{l c c c c c }
\hline \hline \noalign{\smallskip}
 nucl.  & $J^{\rm \pi}$
&\hspace*{1.ex} $T_{\rm 1/2}$\hspace*{1.ex} &\hspace*{1.ex}$Q_{\rm eff}$\hspace*{1.ex} &\hspace*{1.ex} $Q_{\rm theor}$ &\hspace*{1.ex}$Q_{\rm exp}$\\
\noalign{\smallskip}
\hline
\noalign{\smallskip}
${^{55}_{28}}$Ni$_{27}$ & 7/2$^{+}$    & 204.7 ms   &  0.11 &  0.26  & --\\
${^{57}_{28}}$Ni$_{29}$ & 3/2$^{-}$    & 35.6 h     & -0.07 & -0.17  & --\\
${^{77}_{28}}$Ni$_{49}$ & (9/2)$^{+}$  & 128 ms     &  0.16 &  0.20  & --\\
${^{79}_{29}}$Ni$_{50}$ & (5/2$^{+}$)  & 635 ns     & -0.17 & -0.12  & --\\
${^{101}_{50}}$Sn$_{51}$ & (5/2)$^{+}$ & 1.7 s      & -0.13 & -0.21  & --\\
${^{131}_{50}}$Sn$_{81}$ & (3/2$^{+}$) & 56 s       &  0.10 &  0.10  & -0.04(8)\\
${^{133}_{50}}$Sn$_{83}$ & 7/2$^{-}$   & 1.46 s     & -0.23 & -0.17  & --\\
${^{207}_{82}}$Pb$_{125}$ & (1/2)$^{-}$& stable     &  0    &  0     & --\\
${^{209}_{82}}$Pb$_{127}$ & (9/2)$^{+}$& 3.253 h    & -0.29 & -0.26  & -0.3(2)\\
\\
\hline \hline
\end{tabular}
\end{center}
\end{table}

\begin{table}[tH]
\label{tablqoddp}
\caption{Quadrupole moments \textit{Q} (e b) of odd-proton near-magic nuclei.}
\begin{center}
\begin{tabular}{l c c c c c }
\hline \hline \noalign{\smallskip}
 nucl.  & $J^{\rm \pi}$
&\hspace*{1.ex} $T_{\rm 1/2}$\hspace*{1.ex} &\hspace*{1.ex}$Q_{\rm eff}$\hspace*{1.ex} &\hspace*{1.ex} $Q_{\rm theor}$ &\hspace*{1.ex}$Q_{\rm exp}$\\
\noalign{\smallskip}
\hline
\noalign{\smallskip}
${^{55}_{27}}$Co$_{28}$ & 7/2$^{-}$    & 17.53 h    &  0.22 &  0.31  & --\\
${^{57}_{29}}$Cu$_{28}$ & 3/2$^{-}$    & 196.3 ms   & -0.15 & -0.20  & --\\
${^{79}_{29}}$Cu$_{50}$ & (3/2$^{-}$)  & 188 ms     & -0.14 & -0.13  & --\\
${^{99}_{49}}$In$_{50}$ & (9/2)$^{+}$  & 3 s        &  0.35 &  0.35  & --\\
${^{131}_{49}}$In$_{82}$ & (9/2)$^{+}$ & 0.28 s     &  0.40 &  0.28  & --\\
${^{133}_{51}}$Sb$_{82}$ & (7/2)$^{+}$ & 2.34 m     & -0.34 & -0.23  & --\\
${^{207}_{81}}$Tl$_{126}$ & 1/2$^{+}$  & 4.77 m     &  0    &  0     & --\\
${^{209}_{83}}$Bi$_{126}$ & (9/2)$^{-}$& stable     & -0.51 & -0.34  & -0.37(3), -0.55(1)\\
 & & & & & -0.77(1), -0.40(5)\\
\\
\hline \hline
\end{tabular}
\end{center}
\end{table}

%

\subsection{Odd-odd near-magic nuclei}
Using Eqs.(3),(4) and  the values of quadrupole moments of  odd nuclei presented in Tables 2 and 3, we have calculated 
quadrupole moments of corresponding odd-odd nuclei  with both the phenomenological values 
$e{^p_{eff}}=2$, $e{^n_{eff}}=1$ (column $Q_{eff})$  and  the self-consistently calculated quadrupole moments of odd nuclei (column $Q_{theor})$, see Table 4. Unfortunately, the experimental data are scarce  and may be  not achievable, so the $Q_{theor}$ results in Table 4 are 
our predictions of the quadrupole moments in  the exotic unstable nuclei under consideration.  

\begin{table}[h]
\label{selfodd-odd}
\caption{Quadrupole moments \textit{Q} (e b) of odd-odd near-magic nuclei.}
\begin{center}
\begin{tabular}{l c c c c c }
\hline \hline \noalign{\smallskip}
 nucl.  & $J^{\rm \pi}$
&\hspace*{1.ex} $T_{\rm 1/2}$\hspace*{1.ex} &\hspace*{1.ex}$Q_{\rm eff}$\hspace*{1.ex} &\hspace*{1.ex} $Q_{\rm theor}$ &\hspace*{1.ex}$Q_{\rm exp}$\\
\noalign{\smallskip}
\hline
\noalign{\smallskip}
${^{54}_{27}}$Co$_{27}$ & 0$^{+}$      & 193.28 ms  &  --   & --     & --\\
${^{56}_{27}}$Co$_{29}$ & 4$^{+}$      & 77.236 d   &  0.19 &  0.30  & +0.25(9)\\
${^{56}_{29}}$Cu$_{27}$ & (4$^{+}$)    & 93 ms      &  0.14 &  0.28  & --\\
${^{58}_{29}}$Cu$_{29}$ &  1$^{+}$     & 3.204 s    &  0.09 &  0.15  & --\\
${^{78}_{29}}$Cu$_{49}$ & (3$^{-}$)    & 637 s      & -0.18 & -0.21  & --\\
                        & (4$^{-}$)    &            &  4$\times 10^{-5}$ &  -0.03  & --\\
${^{100}_{49}}$In$_{51}$& (6$^{+}$)    & 5.9 s      &  0.24 &  0.21  & --\\
${^{130}_{49}}$In$_{81}$&  1$^{-}$     & 0.29 s       & -0.08 & -0.07  & --\\
${^{132}_{49}}$In$_{83}$& (7$^{-}$)    & 0.207 s    & -0.40 & -0.29  & --\\
${^{132}_{51}}$Sb$_{81}$& (4)$^{+}$    & 2.79 m     & -0.30 & -0.22  & --\\
${^{134}_{51}}$Sb$_{83}$& (0$^{-}$)    & 0.78 s     &  --   & --     & --\\
${^{206}_{81}}$Tl$_{125}$& 0$^{-}$     & --         & --    & --     & --\\
${^{208}_{81}}$Tl$_{127}$& 5$^{+}$     & 3.053 m    & -0.30 & -0.27  & --\\
${^{208}_{83}}$Bi$_{125}$& 5$^{+}$     & 3.68E+5 y  & -0.51 & -0.35  & -0.64(6)\\
${^{210}_{83}}$Bi$_{127}$& 1$^{-}$     & 5.012 d    &  0.21 &  0.16  & +0.136(1)\\
\\
\hline \hline
\end{tabular}
\end{center}
\end{table}

\section{Conclusion}
\label{conclusions}
A reasonable agreement with all available experimental data for quadrupole moments in odd and odd-odd 
near-magic nuclei   has been obtained in our self-consistent calculations using the approximation of no  interaction between odd particles.

The self-consistent approach has a high predictive power required to describe properties of nuclei where there is no experimental data. In this connection, our predictions  for quadrupole moments of nuclei near unstable  $^{56,78}$Ni, $^{100,132}$Sn are of special interest.

\begin{acknowledgement}
The authors thank E.E. Saperstein and V.I. Isakov for useful discussions. 
The work was partly supported by the DFG
and RFBR Grants Nos.436RUS113/994/0-1 and 09-02-91352NNIO-a, by the
Grants NSh-7235.2010.2  and 2.1.1/4540 of the Russian Ministry for
Science and Education, and by the RFBR grants 11-02-00467-a and
12-02-00955-a.

\end{acknowledgement}

\end{document}